\documentclass[12pt]{article}
\usepackage{epsfig}
\newcommand{\tr}{{\rm Tr}}
\begin{document}

\title{\hfill\parbox{3cm}{\normalsize\raggedleft DESY 01-198\\ hep-ph/0111306}\\[30pt]
The Small-$x$ Behaviour of the Singlet Polarized Structure
Function $g_2$ in the Double Logarithmic Approximation }
\author{J. Bartels\footnote{supported by the EU
  TMR-Network `QCD and the Deep Structure of Elementary Particles',
  contract number FMRX-CT98-0194 (DG 12-MIHT).}\\
II.Institut f. Theoretische Physik, Universit\"at Hamburg, Germany\\ \\ 
M.G. Ryskin\footnote{Work supported by the
NATO Collaborate Linkage Grant SA (PST.CLG.976453)5437
and by the RFFI grant 00-15-96610} \\ Petersburg
Nuclear Physics Institute\\ Gatchina, S.-Petersburg 188300, Russia}
\maketitle

\begin{abstract}
The small-$x$ behavior of the singlet contributions to the polarized structure
function $g_2(x,Q^2)$
is calculated in the double-logarithmic approximation of perturbative QCD.
The dominant contribution is
due to the gluons which, in contrast to the unpolarized case, mix with
the fermions also in the small-$x$ domain.
We find a power-like growth in $1/x$ in the
odd-signature parts of the amplitude with the same power as in the
singlet function $g_1(x,Q^2)$ at $x\ll1$.
\end{abstract}

\newpage
\section{Introduction}

The investigation of the structure functions $g_1$ and $g_2$ provides
the basis for the theoretical description  of polarization effects in
deep inelastic lepton nucleon scattering. The $Q^2$-evolution of the
spin dependent function $g_1$ is well known: at $x\sim0(1)$ one can use
the original Altarelli--Parisi equations \cite{AP}, whereas in the region of
very small $x$ the double logarithmic
approximation \cite{BER1,BER2} has to be used and predicts a stronger growth
in $1/x$ than the Altarelli-Parisi equations.
The situation with the other polarized structure function, $g_2$, is more
complicated. Compared to $g_1$, several new features appear \cite{BR}.

First, there exists a twist-3 contribution to $g_2$, $g^{(3)}_2$, which is
not suppressed even for large $Q^2$ \cite{HM}. Part of the function $g_2$
may be obtained from the known structure function $g_1(x,Q^2)$ with the
help of the Wandzura--Wilczek relation \cite{WW}, and from another integral
relation \cite{BK} which relates the twist-3 piece of $g_2$ to the
$(m^2/Q^2$-suppressed) twist-3 corrections to $g_1$. However these
relations are not enough to restore the whole spin dependent function
$g_2(x,Q^2)$. This can be seen from the explicit one loop calculation of
$g_2(x,Q^2)$ on a quark target \cite{ALNR,HW} \footnote{The one loop
function $g_2(x,Q^2)$ on the gluon target was calculated in \cite{GR}.}
and also from the small-$x$ behaviour of the nonsinglet
component of $g_2(x,Q^2)$, calculated in the double logarithmic
approximation (DLA) in \cite{BR}.

Another distinct feature and difference from the
standard leading-twist evolution equations is the fact that the number of
operators which contribute to the twist-3 component of $g_2(x,Q^2)$,
$g^{(3)}_2$, is not fixed but increases
with the moment index $n$ (see, e.g. \cite{BKL,KYU}). Nevertheless, in
the two limits:\\
i) $n\to\infty$ (where $n$ denote the $(x)$-moments)\\
ii) number of colours $N_c\to\infty$\\
it has been shown that the quark-gluon operators decouple from the
quark operator evolution equation \cite{ABH}, and the asymptotic
behaviour of $g_2(x,Q^2)$ in the region $1-x\ll1$ has been derived
\footnote{See also \cite{Sak}for a  recent study of the
twist-3 $(g^{(3)}_2)$ evolution in the $N_c\to\infty$ limit and
\cite {BrK} for a possibility to study the scale dependence of
$g_2(x,Q^2)$ beyond the $N_c\to\infty$ limit.}.

Next, in the region of small $x$ $g_2$ has to be decomposed into pieces
with different signatures.
Whereas unpolarized structure functions and the polarized structure function
$g_1$ have a definite signature
(e.g., $g_1$ is the energy discontinuity of an odd-signature scattering
amplitude), $g_2$ contains both even and odd signature
\cite{IKL,RLH}, and an important part of the small-$x$ analysis of $g_2$
is the separation of the two signature structures. Because of the different
contents of signature, $g_1$ and $g_2$ are expected to have a different
behaviour near $x=0$ (i.e. different powers of $1/x$; see, for example,
\cite{IKL,KL}).

Finally, in the region of small $x$ another new feature arises. It is well
known that the strong ordering of transverse momenta is violated
\cite{BER1,KL,GGFL,EMR}. Instead of $k^2_{i,t}\gg k^2_{i-1,t}$ one has
$k^2_{i,t}\gg\frac{x_i}{x_{i-1}}k^2_{i-1,t}$ (with $x_i\ll x_{i-1}$).
As a result, double log contributions of the form
$(\alpha_s\ln^21/x)^n$ appear which cannot be summed up in the
framework of the conventional $\log Q^2$ evolution. This feature holds for
both $g_1$ and $g_2$, and for $g_1$ it has been discussed in detail in \cite{BER1,BER2}. In particular,
the infrared evolution equation (IREE) \cite{KL} has been used to sum up
all the leading pQCD double logarithms (DL) of the form
$\alpha^n_s\ln^k1/x\ln^mQ^2$ with $m\le n$ and $k+m=2n-1$. In \cite{BR}
similar techniques have been used to investigate the small-$x$ behaviour
of the polarized structure function  $g_2$: so far only the nonsinglet case
has been analyzed. In this paper we present the study of the singlet
part of $g_2$, and we find its asymptotic behaviour at $x\to0$.
In analogy with our study of the nonsinglet part of $g_2$ we consider
the scattering amplitude of a virtual photon an a quark. We will not (yet)
address the question
how our calculations can be used for the polarized structure function
$g_2$ of the proton, i.e. how to set up the initial conditions coming
from the confinement region.

In \cite{BR} we have shown that, for the even signature part of the nonsinglet
$g_2$, the only double logarithmic contributions come from the ladder-type
Feynman diagrams with $t$-channel intermediate states consisting of two quarks
only.
For the odd-signature part of the amplitude one has to consider also
non-ladder graphs where an additional 'soft' $t$-channel gluon ``embraces''
a part of the ladder diagram, in which the transverse momenta of all
partons (gluons and quarks) are larger than the transverse momentum
of this ``embracing'' soft gluon.
The DL contribution of such a graph can be summed up
by the IREE, using the method proposed in \cite{KL}. In ~\cite{BR}
we have analyzed all one and two loop diagrams, contributing to the
nonsinglet component of $g_2$, and we extracted the double logarithms.
In the singlet part of $g_2$, whose analysis is the goal of the present paper,
there are only two new 2-loop diagrams, shown in Fig.1.
Fig.1a has the same ladder structure but with the two-gluon $t$-channel
intermediate state while the crossed box diagram of Fig.1b, as usual,
looses the logarithm. The analysis of these graphs, together with the results
of ~\cite{BR} will enable us to find the small-$x$ behavior of the singlet
$g_2$ structure function.

A comment is in place about the contribution of t-channel states with more
than two gluons. For unpolarized structure functions at very small $x$,
diagrams with any number
of t-channel gluons contribute to the small-$x$ behavior of the scattering
amplitude $\sim 1/x$ (modulo powers of $\ln 1/x$). For the polarized case
(both $g_1$ and $g_2$) the antisymmetric tensor structure (see below) reqires,
for the two gluon state, the polarization of the t-channel gluons to be
different from each other; as a result, the amlitude behaves as $ \sim x^0$
(modulo powers of $\ln 1/x$). Consequently, even at small $x$, quark and
gluon t-chanel states mix, quite in contrast to the unpolarized case.
However, also in the polarized case, there exist t-channel states of gluons
which lead to a small-$x$ behavior $\sim 1/x$. The simplest one
~\cite{Lipatov} consists of
three gluons, has positive C-parity, and is dual to the Odderon solution
discussed in ~\cite{BLV}. This three gluon  contribution, however, is
beyond the DL approximation used in this paper. First, the coupling of this
state to the virtual photon is
suppressed by a few extra powers of $\alpha_s$ which are not accompanied by
the maximal number of logarithms. Next, the contribution of this state to
$g_2$ is suppressed by an extra power of $1/Q^2$, needed to compensate the
dimension of the transverse momenta $k_{1t}$, $k_{2t}$ (of the gluons)
which saturate the antisymmetric tensor $\varepsilon
^{\mu\nu\alpha\beta}$, and reflecting the orbital momentum in the three gluon
state. For these reasons, contributions from diagrams with more than two
gluons will not be considered in this paper.

The outline of the paper is as follows.
In Sect.2 we consider the kinematics of the process and outline our strategy.
In Sect.3 we calculate the DL
small-$x$ contribution of the simplest singlet diagrams shown in Fig.1.
It turns out that the DL asymptotics of Fig.1 is proportional to
the singlet structure function $g_1(x,Q^2)$. Based on this general
property we consider in Sections 3 and 4 the small-$x$ behaviour of the spin
dependent function $g_2(x,Q^2)$. Sections 5 and 6 contain the evolution
equations and their solutions. In a concluding section we present
a brief summary.

\section{Kinematics and definitions}

Let us begin with a brief summary of previous results and with an outline
of our future strategy.
The spin dependent part of the hadronic tensor $iW^A_{\mu\nu}$ of the
DIS lepton-quark scattering amplitude has the form
\begin{equation} \label{1}
T^A_{\mu\nu}=\ \frac M{(pq)}i\varepsilon^{\mu\nu\alpha\beta} \left[
q_\alpha s_\beta T_1(x,Q^2)+q_\alpha\left(s_\beta-\frac{(sq)}{(pq)}
p_\beta\right)T_2(x,Q^2)\right] ,
\end{equation}
where $M,p_\mu$ and $s_\mu$ denote the mass, the four momentum and the
spin (pseudo)vector of the target, respectively. As usual, $(sp)=0$ and
$s^2=-1$, and the spin-dependent structure functions are defined as the
discontinuities of $T_1$ and $T_2$:
\begin{equation} \label{2}
g_{1,2}\ =\ -\ \frac1{2\pi}\mbox{ Im }T_{1,2}\ .
\end{equation}
Throughout this paper we work in the Feynman gauge and use the Sudakov
representation of the momenta of quarks and gluons:
\begin{equation} \label{3}
k_i\ =\ -\alpha_i q'+\beta_ip+k_{ti}\ .
\end{equation}
The photon four momentum is written as $q=q'-xp$, and $p$ and $q'$ are the
two light-like reference vectors. The Jacobian can be written as:
\begin{equation} \label{4}
d^4k_i\ =\ \frac S2\ d\alpha_id\beta_id^2k_{ti}\ .
\end{equation}
Here $S=2q'p$ denotes the center of mass energy, and we assume that the quark
target mass is small: $M^2\ll Q^2=-q^2$. So
one may neglect the value of this mass $M$ everywhere, except for the
mass term in the target density matrix. This mass term is needed in order
to obtain a nonzero spin trace proportional to the first power of $M$
(see Eq.({\ref{1}))
%%%%%%%%%%
\footnote{Hence we may put $p^2\simeq p'^2=0$.}
%%%%%%%%%%%%
. Finally, we recall from ~\cite{BR} that, in
order to obtain in (\ref{1}) a nonzero contribution to $g_2$, we need a
transverse polarization of the target ($s_{\mu}=s_{\mu t}$): longitudinal
polarization of the target $s_\mu=p'_\mu/m$ contributes to the structure
function $g_1$ only, while the transverse $(s_\mu=s_{t\mu})$ polarization
gives the sum $g_\perp=g_1+g_2$.
\begin{figure} 
\begin{center}
\epsfig{file=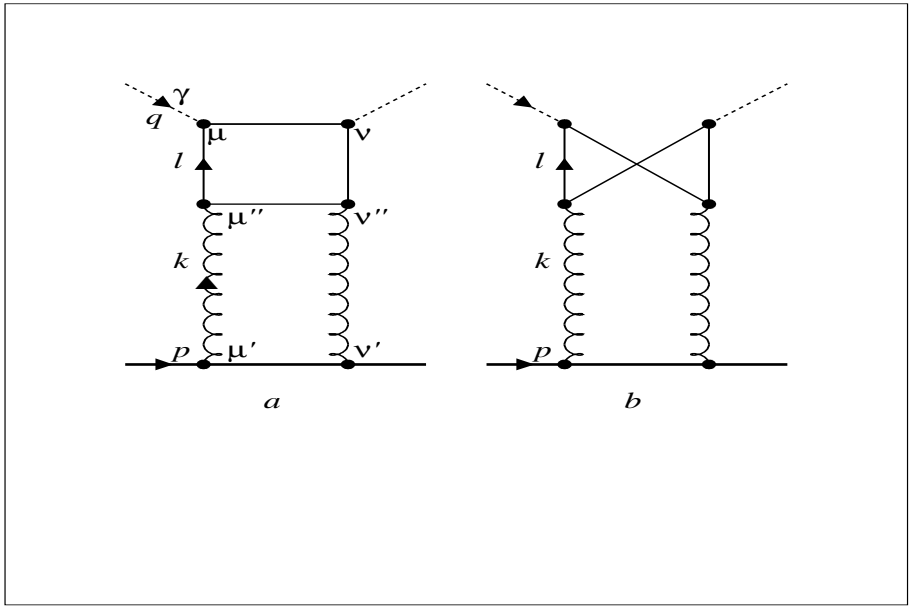,width=12cm,height=8cm}
\caption{\sl The lowest order Feynman diagrams which contain a gluon
t-channel state}
\end{center}
\label{generic}
\end{figure}

Let us briefly recapitulate our previous analysis of the polarized structure
function. Consider a generic ladder diagram (Fig.2) and begin at the bottom.
Since the spin dependent part of the hadronic tensor $W_{\mu\nu}$
is antisymmetric (proportional to the $\varepsilon$-tensor),
in the target density matrix we have to start from the $\gamma_5\hat s$ part:
\begin{equation} \label{5}
\frac12\ (\hat p+m)(1-\gamma_5\hat s)\ .
\end{equation}
When calculating the trace, this $\gamma_5$-matrix gives
the antisymmetric tensor $\varepsilon^{\mu\nu\alpha\beta}$. As we have said
before, for the present analysis we chose transverse polarization of the
target quark.

Moving upwards inside the ladder diagrams, we encounter t-channel
intermediate states which consist of an quark-antiquark pair or of two gluons.
For the quark $t$-channel state, we have found in \cite{BR} two different
polarization structures which we will denote by $'s'$ and $'k'$:
\begin{eqnarray}
&& 's'\ \Rightarrow\ \gamma_5\hat s_t\ ,
\label{8} \\
&& 'k'\ \Rightarrow\ (2sk)\cdot\gamma_5\hat k\ ,
\label{9}
\end{eqnarray}
where $k_\mu$ is the $t$-channel quark momentum. At the target we begin with
the `s' structure, but at any gluon rung above we have both possibilities:
either to continue with `s' or start a `k branch'. The latter continues with
`k' structures only, until at the upper end of the diagram it ends with the
coupling to the photon. There is no return from a `k branch' to an `s branch'.
Morover, the `k branch' contains only odd signature, and its evolution is the
same as in $g_1$, whereas the `s branch' contains both even and odd signature
and has slightly modified evolution equations. Nonladder structures
cancel for even signature branches, whereas for odd signature they lead to an
additional term in the evolution equation.

What is new in the present singlet case, is the two-gluon intermediate state.
We first note that this intermediate state must have the same
antisymmetric structure (with respect to the gluon polarizations $\mu'$
and $\nu'$) as the $\varepsilon$ polarization tensor. Thus we cannot assign
to both gluons the same longitudinal
polarizations which would lead to the small-$x$ behavior $\sim 1/x$.
The best one can do for retaining the largest
power of $1/x$ is to assign the longitudinal
polarization $(e_{\mu'}\propto q'_{\mu'})$ to one of the two gluons and
the transverse polarization $(e_{\nu'}=e_{\nu't})$ to the other gluon.
Therefore, as mentioned before, in contrast to the unpolarized structure
function $f_1(x,Q^2)\sim1/x$, the prediction for the small-$x$ behaviour of the
polarized distributions goes as
\begin{equation} \label{5a}
g_{1,2}(x,Q^2)\sim\mbox{ const }\cdot F(\alpha_s\ln Q^2\ln1/x,
\alpha_s\ln^21/x)\ ,
\end{equation}
where the function $F$ denotes the leading (DL) corrections to the
asymptotics of the Born singlet spin
dependent structure function: $g_{1,2}(x,Q^2)\sim\,const$ as $x\to 0$.
Speaking in terms of the complex
angular momentum $j$ plane, both the two $t$-channel quark and two gluon
exchange lead to the rightmost singularity at $j=\omega=0$, and even at very
small $x$ for the singlet spin dependent structure functions we have to
account for the transitions between the quark and gluon $t$ channel
states. Continuing with the spin structure of the $t$-channel two gluon
state we
have to recall that the longitudinal gluon polarization $e_{\mu'}$
is contained in the term $q'_{\mu'}p'_{\mu''}/(q'p')$ in
the spin part of gluon propagator
\begin{equation} \label{6}
d_{\mu'\mu''}\ =\ g_{\mu'\mu''}\ =\ g^\perp_{\mu'\mu''} +
\frac{p'_{\mu'}q'_{\mu''}+q'_{\mu'}p'_{\mu''}}{(p'q')}\ ,
\end{equation}
where the vector $q'_{\mu'}$ ($p'_{\mu''}$) is contracted with the
spin structure at the lower (upper) end of the
$t$-channel gluon. The transverse polarization is contained in
$\ g^\perp_{\mu'\mu''}$.
At first sight there are two antisymmetric spin structures with one
longitudinal gluon:
\begin{equation} \label{7}
\mbox{(a)}\;\;\;\;\;\;\;\;\; \varepsilon^{\mu\nu\alpha\beta}
s^\perp_\alpha p'_\beta
%\label{7}
\end{equation}
and
$$ \mbox{(b)}\;\;\;\;\;\;\;\; s^\perp_\mu p'_\nu-p'_\mu s^\perp_\nu\ .$$
%\begin{eqnarray}
%\mbox{(a)} && \varepsilon^{\mu\nu\alpha\beta} s^\perp_\alpha p'_\beta
%\nonumber \\
%&& \hspace*{-2cm} \mbox{and} \\
%\label{7}
%\mbox{(b)} && s^\perp_\mu p'_\nu-p'_\mu s^\perp_\nu\ . \nonumber
%\end{eqnarray}
However, the last structure has unnatural parity ($s_\mu$ is a
pseudovector) and never contributes to our amplitude Eq.(\ref{1}).
So it is enough to consider the first structure, Eq.(\ref{7}), only.
Since this structure is the same as in the case of $g_1$, we can make use
the results of ~\cite{BER2}.

It will be the central task of this paper to compute the evolution kernels
of the different t-channel states and the transitions from one to another.
Altogether, we have three different states: the two spin structures of
quark-antiquark states, (6) and (7), and the two gluon state.
As usual, the DL-contribution comes from the kinematical region of
$$
1\ \gg\ \beta_1\ \gg\ \beta_2\ \gg\ \ldots\ >\ x
$$
and
\begin{equation} \label{10}
\ldots\ \ll\ \alpha_1\ \ll\ \alpha_2\ \ll\ \ldots\ \ll\ 1\ .
\end{equation}
The contour of the $\alpha_i$ loop integration will be closed around the pole
of the propagator of the $i$-th $s$-channel (horizontal line)
particle. Thus all the $s$-channel particles lie on mass-shell,
and the $t$-channel propagators can be approximated
by $1/k^2_i\simeq/k^2_{it}$, as $\alpha_i\beta_is\ll
k^2_{it}$. So the $i$-th loop integral takes the form
\begin{equation} \label{11}
C\ \frac{\alpha_s}{2\pi}\ \frac{d\beta_i}{\beta_i}\ \frac{dk^2_{it}}{
k^4_{it}}\cdot numerator\ .
\end{equation}
The logarithmic $\beta_i$ integration is written
explicitly, while, in order to obtain the second logarithm (from the integral
over $dk^2_{it}$), we need to select the term proportional to $k^2_{it}$
in the numerator, coming from the spin part of the diagram.
As discussed before, the gluon$\to$gluon case we take from ~\cite{BER2};
the evolution of the `s branch' and `k branch' quark structures were
obtained in ~\cite{BR}, the transition from `s branch' to `k branch' is also
contained in ~\cite{BR}, and the transition from `k branch' to `s branch'
vanishes. Collecting these results in terms of a matrix whose columns (rows)
are labelled by the spins structure of the t-channel states below (above) 
the transition, we have for the $numerator$ in (\ref{11}):
\begin{equation}
\left( \begin{array}{ccc}
            gg & kg & sg \\
            gk & kk & sk \\
            gs & ks & ss \end{array} \right)
          = \left( \begin{array}{ccc}
            4 & -1 & ? \\
            2 & 1  & 0 \\
            ? & -1 & 1 \end{array} \right)
\end{equation}
Here the labels ($g$, $k$, $s$) refer to  the spin structures of the gluon,
the `k' structure of the quark, and the `s' structure of the quark, resp.
The 2x2 block matrix in the upper left corner coincides with $g_1^s$ and
has been taken from ~\cite{BER2}; the 2x2 block matrix in the lower left
corner, on the other hand, is the same as in the nonsinglet $g_2$
case and has been derived in ~\cite{BR}. What needs still to be computed are
the matrix elements denoted by `$?$': in the two following
sections we will show that both elements vanish.

The color factors $C$ are the same as in the unpolarized scattering: for the
gluon$\to$gluon it is equal to
$C_A=N_c$ ($N_c=3$  is the number of colours), for
the quark $\to$ gluon and quark $\to$ quark splitting we have
$C_F=(N^2_c-1)/2N_c$,
and for the gluon $\to$ quark transition the color factor is
$2T_F=n_F$ ($n_F$ is the number of light quark flavours).
Combining these color factors with the elements of the $numerators$ matrix
in eq.(13), we have all ingredients to the infrared evolution equations.

\begin{figure}
\begin{center}
\epsfig{file=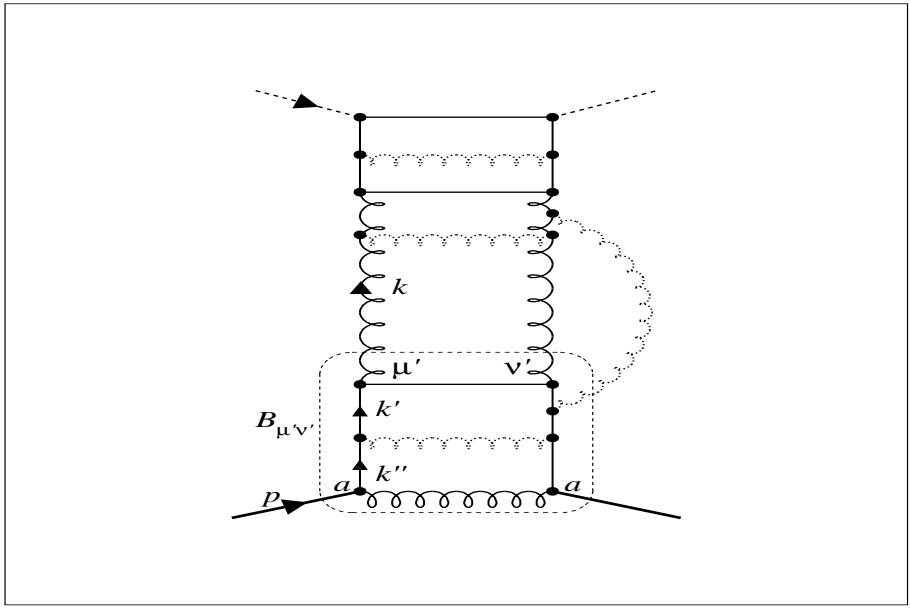,width=12cm,height=8cm}
\caption{\sl A generic Feynman diagram for the scattering of a
virtual photon on a quark target}

\end{center}
\label{twoloop}
\end{figure}

\section{Singlet two-loop diagram}

We start with the amplitude Fig.1a and use the transverse vector $s_{t\mu}$
The spin dependent lower part of Fig.1a takes the form:
\begin{equation} \label{12}
B_{\mu'\nu'}=\ -\frac12\mbox{ Tr }\left[\gamma_{\nu'}(\hat p+m)\gamma_5
\hat s\gamma_{\mu'}(\hat p-\hat k+m)\right]=\ -2i
\varepsilon^{\mu'\nu'\alpha\beta}s^\perp_\alpha k_\beta\ ,
\end{equation}
where in order to obtain the leading (in $1/x$) contribution we have to
keep the longitudinal component of the gluon momentum
$k_\beta\simeq\beta p'_\beta$.

Next, to simplify the calculation of the trace of the upper quark loop we
make  use of gauge invariance. The amplitude
$U^{\mu\nu}_{\mu''\nu''}$, which corresponds to the upper block and
includes both the box (Fig.1a) and crossed box (Fig.1b) diagrams,
satisfies the properties
\begin{equation} \label{13}
U^{\mu\nu}_{\mu''\nu''}\cdot q_\mu\ =\ 0
\end{equation}
and
\begin{equation}\label{14}
U^{\mu\nu}_{\mu''\nu''}k_{\mu''}\ =\ U^{\mu\nu}_{\mu''\nu''} (k_t+\beta
p'-\alpha q')_{\mu''}\ =\ 0\ .
\end{equation}
Based on the gluon gauge invariance (\ref{14}) one may replace the
momentum $p'_{\mu''}$ in the longitudinal gluon polarization vector
(i.e. spin part of the gluon propagator (\ref{6})) by 
$-k_{t\mu''}/\beta$; the  contribution of the term $-\alpha q'_{\mu''}$
does not contain the double logarithm and may be neglected within the
kinematical domain (\ref{10}). We contract the tensor in (14) with the
gluon numerators (9) and make use of this replacement.
The polarization of the $t$-channel gluons is then described by the
tensor structure:
$$
T_{\mu''\nu''}\ \propto\ \frac{\tilde e_{\mu''}k_{t\nu''}-k_{t\mu''}
\tilde e_{\nu''}}\beta\ =\ \frac{(ks^\perp)}\beta
\varepsilon^\perp_{\mu''\nu''}=\ \frac{(ks^\perp)}{(p'q')\beta}
\varepsilon_{\mu''\nu''\alpha\beta}p'_\alpha q'_\beta\ ,
$$
where $\varepsilon_{\mu\nu}^\perp$ is the two dimensional antisymmetric
tensor, and the polarization vector has the form $\tilde e_{\mu''}=
\varepsilon_{\mu''\alpha}^{\perp} s^\perp_\alpha$. In this way the expression
(\ref{12}) can be rewritten as
\begin{equation} \label{15}
B_{\mu''\nu''}\simeq\ 2i\beta\varepsilon^{\mu''\nu''\alpha\beta}
p'_\alpha s^\perp_\beta\ \Rightarrow\ 2i\,\frac{(ks^\perp)}{(p'q')}\
\varepsilon^{\mu''\nu''\alpha\beta} p'_\alpha q'_\beta\ .
\end{equation}
Since the transverse spin vector $s_{t\mu}$ corresponds to the
spin-flip amplitude from the photon side we choose one transverse and
one  longitudinal photon polarization vector, $E_\mu=E^\perp_\mu$ and
$E_\nu=E^{\|}_\nu=\frac1{\sqrt{Q^2}}(q'+xp')$. With the help of the photon
gauge invariance (\ref{13}) the last vector can be replaced by
$E^{\|}_\nu=2xp'/\sqrt{Q^2}$. Thus, finally, in the case of Fig.1a we have
to calculate the trace
\begin{equation} \label{16}
{\rm Tr}^a\ =\ \frac{2i(ks^\perp)}{(p'q')}\ \varepsilon^{\mu''\nu''\alpha\beta}
p'_\alpha q'_\beta\left[\gamma_\nu''\hat\ell\widehat E^{\|}(\hat
q+\hat\ell)\widehat E^\perp\hat\ell\gamma_\mu''(\hat\ell-\hat k)\right].
\end{equation}
Using the identity
\begin{equation} \label{17a}
\gamma_\nu\hat k\gamma_\mu=-i\varepsilon_{\mu\nu
\lambda\sigma}k_\lambda\gamma_5\gamma_\sigma+\mbox{ (the term symmetric
under $\mu\rightleftharpoons\nu)$}
\end{equation}
 we obtain
\begin{eqnarray}
{\rm Tr}^a &=& \frac{4(ks^\perp)}{(p'q')}\bigg\{(p'(\ell-k)){\rm
Tr}\left[\gamma_5\hat q'\hat\ell\hat E^{\|}(\hat q+\hat\ell)\hat
E^\perp\hat\ell\right]\ - \nonumber\\
\label{18}
&-&(q'(\ell-k)){\rm Tr}\left[\gamma_5\hat p'\hat\ell\hat E^{\|}(\hat
q+\hat\ell)\hat E^\perp\hat\ell\right]\bigg\}.
\end{eqnarray}
Putting in (\ref{18}) $E^{\|}=2xp'/\sqrt{Q^2}$, it is easy to see that
the largest (and the only DL) contribution (where $q'$ is multiplied by
the momentum $k$, and $p'$ by momentum $\ell'$; note that
$\alpha_\ell\gg\alpha_k$ and $\beta_k\gg\beta_\ell$) comes from the
second term and reads:
\begin{equation} \label{19}
{\rm Tr}^a\ \approx\ \frac{4(ks^\perp)}{(p'q')}\ (q'k)(2p'\ell){\rm Tr}
\left[\gamma_5\hat E^{\|}\hat q\hat E^\perp\hat\ell\right].
\end{equation}
As the $s$-channel quark with momentum $(k-\ell)$ is on mass shell, the
product $(q'k)(2p'\ell)/(q'p')=-\beta_k\alpha_\ell
s\simeq (k-\ell)^2_t$. So after the azimuthal angular integration we
get
$$
4(ks^\perp)(k-\ell)^2_t\ell_\mu\ =\ 2k^2_t\ell^2_ts^\perp_\mu\ ,
$$
that is
\begin{equation} \label{20}
{\rm Tr}^a\ \simeq\ 2k^2_t\ell^2_t{\rm Tr}\left[\gamma_5\hat E^\perp
\hat E^{\|}\hat q\hat s^\perp\right]\Longrightarrow\
i\varepsilon^{\mu\nu\alpha\beta} q_\alpha s^\perp_\beta (8k^2_t\ell^2_t)\
,
\end{equation}
where the first factor is exactly the structure we are looking for in
the hadronic tensor (\ref{1}), and the last factor $8k^2_t\ell^2_t$
is needed to keep the leading logarithms in the $dk^2_t$ and
$d\ell^2_t$ integrations, as it was discussed at the end of Sect.2.

Finally, let us consider the trace corresponding to the crossed box
diagram of Fig.1b:
\begin{equation} \label{21}
\tr^b=\ 2i\frac{(ks^\perp)}{(p'q')}\varepsilon^{\mu\nu\alpha\beta}
p'_\alpha q'_\beta\tr\left[\gamma_\nu(\hat q+\hat\ell-\hat k)\hat
E^{\|}(\hat\ell -\hat k)\gamma_\mu\hat\ell\hat E^\perp(\hat
q+\hat\ell)\right].
\end{equation}
The product of the three $\gamma$-matrices $(\hat\ell\hat E^\perp(\hat
q+\hat\ell))$ may be decomposed as
\begin{equation} \label{22}
\hat\ell\hat E^\perp(\hat q+\hat\ell)\ =\ c_\delta\gamma_\delta+
d_\delta\gamma_5\gamma_\delta\ .
\end{equation}
The last term has unnatural parity and may be neglected, while
the vector $c_\delta=\frac14\tr[\hat\ell\hat E^\perp(\hat
q+\hat\ell)\gamma_\delta]$. Using this fact and the identity (\ref{17a})
we get
\begin{eqnarray}
\tr^b &=& \frac{(ks^\perp)}{(p'q')}\bigg\{\tr\left[\hat p'\hat\ell\hat
E^\perp(\hat q+\hat\ell)\right]\cdot\tr\left[\gamma_5\hat q'(\hat
q+\hat\ell-\hat k)\hat E^{\|}(\hat\ell-\hat k)\right]\ - \nonumber\\
\label{23}
&-& \tr\left[\hat q'\hat\ell\hat E^\perp(\hat
q+\hat\ell)\right]\cdot\tr\left[\gamma_5\hat p'(\hat q+\hat\ell-\hat
k)\hat E^{\|}(\hat\ell-\hat k)\right]\bigg\}.
\end{eqnarray}
The second trace in the first term on the rhs of (\ref{23}) vanishes as it
contains only one transverse vector $(\ell-k)_t$, while the last term in
(\ref{23}) is zero since in the second trace $E^{\|}_\mu\propto
p'_\mu$. Therefore we neglect the crossed box contribution (after the
trick based on the gauge invariance was applied) and the whole DL
result, given by the ladder-type Fig.1 diagrams, reads:
\begin{equation} \label{24}
g^s_\perp(x)\ =\ -e^2_q\cdot2c_F\left(\frac{\alpha_s}{2\pi}\right)^2
\int\frac{d\beta_k}{\beta_k}\int\frac{dk^2_t}{k^2_t}\int
\frac{d\ell^2_t}{\ell^2_t}
\end{equation}
(here $e_q$ is the electric charge of the quark). In (26), the limits of
integration follow from (11) and the discussion after (11).

We have to note that, within the DL approximation, this result for
$g_\perp=g_1+g_2$ (\ref{24}) coincides with the lowest order singlet
function $g^s_1$. In other words, to this order of accuracy we obtain
the singlet $g^s_2(x)=0$. This is the consequence of the fact that in
the lower part of Fig.1 graph the Born structure function $g^B_2=0$.
For the one loop approximation (order $\alpha_s$) it is known ~\cite{ALNR}
that at small $x$ the one loop result is given by:
$$
g_2(x,Q^2)\simeq\frac{e^2_qC_F\alpha_s}{2\pi}x\ln\frac{Q^2}{m^2x},
$$
i.e. is much smaller than
$$g_1(x,Q^2)\sim\frac{e^2_qC_F\alpha_s}{2\pi}\ln\frac{Q^2}{m^2x}.$$
So it is only at the two loop level (or above) that
the nonsinglet structure function $g^{n.s.}_2$ becomes comparable to the
value of $g^{n.s.}_1$ at $x\ll1$. Thus one expects a non-zero DL
contribution to the small-$x$ behavior of the singlet
$g_2$ component at earliest at two loops. Our result (26), however, shows
that at the two loop level it still vanishes. We therefore have to turn to
three or more loops.
On the other hand, we know already from the argument given in section 2
that in the DL approximation
the gluon contributions both to $g_\perp=g_1+g_2$ and to $g_1$
are driven by the same spin structure (\ref{7}). Therefore we may expect
that the distributions $g^s_2$ and $g^s_1$ will have the same asymptotic
small-$x$ behavior. In the following section we will show that this is
indeed the case.

\section{Higher order contributions}

We now turn to higher order corrections to Figs.1a and 1b.
As a generic diagram, we consider
Fig.2. Beginning with the quark target at the bottom, we first have
the same evolution as in the nonsinglet case.
After the first $s$-channel gluon the initial quark density matrix
$\frac{1}{2}(\hat{p}+m)(1-\gamma_5\hat s)$ gives
\begin{equation}
\frac12\sum_a\gamma_a(\hat p+m)(-\gamma_5\hat s)\gamma_a=
m\gamma_5\hat s,
\end{equation}
where we omit the terms which do not depend on the quark spin
vector $s_\mu$. In this way we obtain the structure $\gamma_5\hat s$.
At the next step this spin structure, together with the $t$-channel quark
propagators $\hat k^{''}$, produces the second structure
$(2sk)\cdot\gamma_5\hat k$:
\begin{equation}
\hat k^{''}\gamma_5\hat s\hat k^{''}\ =\ (k^{''2})\gamma_5\hat
s-(2sk^{''})\cdot\gamma_5\hat k^{''}.
\end{equation}
Consequently, starting from two s-channel gluons, the quark density
matrix in the upper $(k')$ cell of the block contains the two structures
$\gamma_5\hat{s}$ and $(2s^{\perp} k')\gamma_5 \hat k'$. So our diagram
splits into two branches, and we have to follow each branch seperately.
The latter one, $\gamma_5\hat k'$, is equivalent to the density
matrix of longitudinally polarized quark with the momentum $k'_\mu$. So
this spin dependent part of the heavy photon--quark $k'$ amplitude is
described by the photon--quark structure function $g_1(\gamma^*q)$
\begin{equation} \label{25}
T_{\mu\nu}(\gamma^*q)\ =\ \frac{i\varepsilon^{\mu\nu\alpha\beta}}{
(k'q)}\ q_\alpha\cdot k'_\beta\cdot T'_1(x',Q^2)
\end{equation}
with Im $T'_1=-2\pi g_1(\gamma^*q)$ and $x'=Q^2/2(k'q)$. The integration
over the azimuthal direction of the vector $k'_t$ $(\int
d\varphi(2s^\perp k^\perp)k'_\mu=k^2_ts^\perp_\mu)$ replaces the tensor
$\varepsilon^{\mu\nu\alpha\beta}q_\alpha k'_\beta$ by
$\varepsilon^{\mu\nu\alpha\beta}q_\alpha s^\perp_\beta$ and gives the
factor $k'^2_t$ needed to save the DL structure of the $dk'^2_t$
integral. As a result, the branch in $g_\perp$ which originates from the
`k' quark structure, $2(s^\perp k')\gamma_5\hat k'$, is proportional to
$g^s_1$.
So we can make use of our results ~\cite{BER2} for $g_1^s$: we know the
transitions from
the gluon t-channel to the quark t-channel (and vice versa), and we also know
the latter never contains the `s' spin structure. In other words, we have \
shown that that in (13) the upper right element, $sg$, vanishes.

Following the other branch, which starts from the `s' structure,
$\gamma_5\hat s^\perp$, at the upper end of the lower block, we note
that it does not lead to any double logarithm at all.
To show this we again make use of the gauge
invariance, that is the condition analogous to (\ref{14}) but with
respect to the lower block: $B_{\mu'\nu'}k_{\mu'}=0$. Based on this
condition we replace the vector $q'$ by $-k_{t\mu'}/\alpha_k$. After
this the polarization of the only $t$-channel gluon state, which may
give the DLogs is described by the pure transverse tensor
$T_{\mu'\nu'}\propto \varepsilon^\perp_{\mu'\nu'}\cdot(ks^\perp)/\alpha$.
On the other hand, the $\gamma_5\hat s^\perp$ quark structure produces
the tensor $B_{\mu'\nu'}=-4i\varepsilon^{\mu'\nu'\alpha\beta}
s^\perp_\alpha(k'-k)_\beta$ orthogonal to
$\varepsilon^\perp_{\mu'\nu'}$.
As a result, in DL we have no transition from the `s' structure to gluon
states: in the lower left element, $gs$, vanishes.

In summary, the only DL singlet contributions start from the
$\gamma_5\hat k'$ quark
structure, and the gluon evolution is the same as in $g_1^s$.
It therefore may be written as the convolution of the nonsinglet
structure function $g_2$ with the singlet function $g^s_1$.
Note that we have to use not
the whole DL nonsinglet function $g^{n.s.}_2$, but only its part $g^k_2$
which corresponds to the structure $\gamma_5\hat k'$. This part can be
easily extracted from ref.\cite{BR}, while the DL singlet function
$g^s_1$ needed in the convolution can be taken from \cite{BER2}.

This value of $\omega_s$ is
considerably larger than the value of corresponding "intercept" of the
nonsinglet function $g^{n.s.}_2$, where
$\omega_{n.s.} =\omega_0 =\sqrt{(2C_F\alpha_s)/\pi}$ (the ratio of the two
values, $\omega_s/\omega_0$ is about 2.6). Therefore the asymptotic
behaviour of the spin dependent singlet structure function
$g^s_2(x,Q^2)$ is almost completely driven by the DL behavior of the singlet
$g^s_1(x,Q^2)$ distribution.

Let us return to the signature question. As we have mentioned in section 2,
the analysis of the nonsinglet case in ~\cite{BR} has shown that
the `s' branch of the quark state contains contributions to both signatures,
wheras the `k' branch has negative signature only. Since in our DL
the two gluon state communicates only with the `k' branch, it
connects to negative signature only. The only place where even signature
can appear is the `s' structure of the quark state which is the same as
in the nonsinglet function $g_2^{n.s.}$.

The negative signature in the gluon t-channel state can be seen rather
directly. Indeed, as it was discussed in \cite{BR}, one obtains the negative
signature term keeping the photon (or gluon) momentum in the trace over
the quark loop. In the quark $\to$ gluon transition the trace reads
\begin{equation} \label{26}
\tr^{gq}\ =\ \tr\left[\gamma_5\hat k'\gamma_{\nu'}(\hat k'-\hat
k)\gamma_{\mu'}\right]\ ,
\end{equation}
where the gluon momentum $k$ gives the only nonzero contribution. The
same can be seen in the gluon to quark transition. Recall our
two loop calculation of Sect.3. Here the leading DL contributions comes
from the second term of the rhs of Eq.(\ref{18}). In this term we
have to keep the product $(q'k)$, which includes the gluon momentum $k$
and the photon momentum $\hat q$ in the trace (see (\ref{19})) in order
to obtain the double logarithms.

Let us finally comment on a point which, at first sight, might look
dangerous.  Namely, starting from the spin structure \begin{equation}
\label{27} -\ \frac M{(pk)}\ i\varepsilon^{\mu\nu\alpha\beta} k_\alpha
p_\beta\ \frac{(sk)}{(pk)} \end{equation} in the lower nonsinglet part of
the Fig.2, where the gluon momentum $k_\alpha$ plays the role of the photon
momentum $q$ in Eq.(\ref{1}), it looks as if one might get a much larger
small-$x$ behavior, $g_2 \sim 1/x$, than obtained in the previous
discussions. In comparison with (\ref{12}) this term $2(s^\perp k)\cdot
i\varepsilon^{\mu\nu\alpha\beta}
k_\alpha p_\beta\simeq k^2_t\cdot i\varepsilon^{\mu\nu\alpha\beta}
s_\alpha\cdot p_\beta$ is enhanced by the factor $1/\beta_k$ as the
longitudinal component of gluon momentum
$k_\alpha\simeq\beta_kp'_\alpha$. However an extra factor $1/\beta_k$
destroys the logarithmic structure of the $d\beta_k/\beta_k$ integral.
Even more, we will show that such a contribution vanishes. Indeed, the
integration in Fig.2 over the rapidity of the upper horizontal quark line of
the lower block $d\beta_{k'}/\beta_{k'}$ may be written in terms of the
$x'=k^2/2(pk)\simeq k^2_t/2(pk)$ variable. This last quark with
momentum $(k'-k)$ is on mass shell. So, keeping the transverse momenta
$k'_t$ and $k_t$ fixed, we have
$$
\beta_{k'}\ =\ \frac{|k'-k|^2_t}{\alpha_k S}\ \ =\
\frac{|k'-k|^2_t}{2p'k}
$$
and therefore $d\beta_{k'}/\beta_{k'}=dx'/x'$.

On the other hand,
we invoke the argument which leads to the Burkhardt--Cottingham sum rule
\cite{BC}
\begin{equation} \label{28}
\int\limits^1_0 dx'g_2(x',Q^2)\ =\ 0\ .
\end{equation}
This sum rule can be derived from an unsubtracted disperison relation for
the photon - proton forward scattering amplitude. For the very special
hilicity structure which contributes to $g_2$, this amplitude leads to a
cross section which falls with energy faster than $1/s$; therefore, the
integral $\int ds' g_2(s')$ is finite, and from the large-$s$ behavior of the
dispersion relation one finds (32). The same large-$1/x$ condition applies
to our starting function $g_2^{n.s.}$ (shown in ~\cite{BR}), therefore
we conclude that
\begin{equation}
\int\frac{dx'}{x'}\ g_2(x',k^2)\ \frac{k^2_t}{(pk)}\ = \int
dx'g(x',k^2)\ =\ 0
\end{equation}
Thus the term which looked so dangerous, in fact, does not contribute to
double logarithmic structure function $g_2$.

\section{Infrared evolution equations}

With results of the previous sections we now turn to the infrared evolution
equations (IREE). We are interested in the scattering of a virtual photon on
a transversely polarized quark target: $T^s=T_1+T_2$, and we will
concentrate on the odd signature part. In order to formulate our coupled
evolution equations, we first have to generalize to different target
structures. As outlined in section 2, inside our diagrams we have to
consider two different spin structures in the quark $t$-channel state
(\ref{8}),(\ref{9}), labelled as $q_s$ and $q_k$, and the gluon spin structure
(\ref{7}). Correspondingly we introduce the three component vector:
\begin{equation} \label{3.1}
T\ =\ \left(\begin{array}{l}
T(\gamma g) \\ T(\gamma q_k)\\ T(\gamma q_s) \end{array} \right).
\end{equation}
For our scattering of a virtual photon on transversely polarized quark target
we will be interested in the $q_k$-component.
We write $T$ as a Mellin transform:
\begin{equation}\label{3.2}
T\ =\ \int\limits^{i\infty}_{-i\infty} \frac{d\omega}{2\pi i}\left(
\frac S{\mu^2}\right)^\omega\xi(\omega)R(\omega,y)\ ,
\end{equation}
where $R(\omega,y)$ is a three -component vector, defined on analogy with
(\ref{3.1}), and $\omega=j$ denotes angular momentum. The signature
factor $\xi(\omega)=$ has the form:
\begin{equation} \label{3.3}
= \quad \frac{e^{-i\pi\omega}-1}2\ \approx\ \frac{-i\pi\omega}2\ ,
\end{equation}
and
\begin{equation} \label{3.4}
y\ =\ \ln\left(\frac{Q^2}{\mu^2}\right).
\end{equation}
The structure function $g_{\perp}^s$ is obtained from the discontinuity
(\ref{2}), and we have to take into account both DL-contributions and
$i\pi$-terms. Before we can write down the IREE,
 we have to introduce further
auxiliary quantities: scattering amplitudes $F_{ij}$ (with $i,j=g,q_k,q_s$)
which describe the scattering of quark or gluon with spin structure $i$
on a target with structure $j$ (loosely speaking, these scattering amplitudes
are obtained if, in Fig.2, we remove the
coupling to the photon at the upper end). We write them as Mellin transforms,
i.e. the $F_{ij}$ are partial waves and depend upon angular momentum
$\omega$. By $F_0$ we denote the 3x3 matrix composed of the $F_{ij}$
(the subscript refers to the color singlet quantum number):
\begin{equation} \label{3.6}
F_0\ =\ \left(\begin{array}{lll}
F_{gg} & F_{kg} & F_{sg} \\
F_{gk} & F_{kk} & F_{sk} \\
F_{gs} & F_{ks} & F_{ss} \end{array} \right)
\end{equation}
(it is the exact analogue to $F_0$ in \cite{BER2} and to $f^{(-)}_0$ in
\cite{BER1}).
\begin{figure}
\begin{center}
\epsfig{file=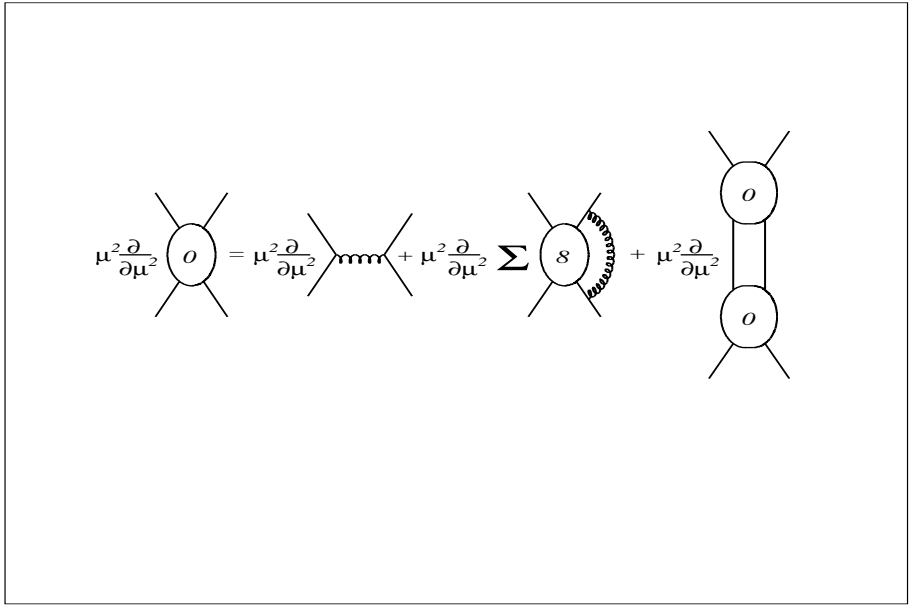,width=12cm,height=8cm}
\end{center}
\caption{\sl Structure of the infrared evolution equations (IREE)}
\label{IREE}
\end{figure}

We now turn to the evolution equations. The IREE (illustrated in Fig.3)
describes the variation of the amplitude
with respect to the infrared cutoff $\mu^2$:
\begin{equation} \label{3.5}
-\mu^2\frac{\partial R}{\partial\mu^2}\ =\
\left(\omega+\frac\partial{\partial y}\right)R\ .
\end{equation}
This differential operator stands on the the left-hand side of the IREE for
$T$, which is illustrated in Fig.3. The right-hand side of the IREE
is derived from the observation that the dependence upon the cutoff $\mu$
resides in the intermediate state with lowest virtuality (Fig.3):
the $\mu$-derivatives
of the amplitudes are equal to $R$ times quark or gluon scattering
amplitudes $F_{ij}$ with the external legs having transverse momenta
close to $\mu$. In terms of the matrix $F_0$, the evolution equation for the
vector $R$ becomes:
\begin{equation} \label{3.7}
\left(\omega+\frac\partial{\partial y}\right)R\ =\ \frac1{8\pi^2}\
F_0R\ .
\end{equation}
The matrix $F_0$ satisfies a nonlinear evolution equation (Fig.3):
\begin{equation} \label{3.8}
F_0(\omega)\ =\ \frac{g^2}\omega\ M_0-\frac{g^2}{2\pi^2\omega^2}\
G_0F_8(\omega)+\frac1{8\pi^2\omega}\ F_0(\omega)^2.
\end{equation}
Here the matrix $M_0$ contains the evolution kernels (13) which we have
computed in the previous section or collected from earlier studies:
\begin{equation} \label{2.8}
M_0\ =\ \left(\begin{array}{ccc}
4C_A & -2T_F & 0\\
2C_F & C_F & 0\\
0 & -C_F & C_F \end{array} \right)
\end{equation}
As we have discussed before, for the
gluon state (upper left corner) the infrared evolution is described by the
same splitting
function as the singlet structure function $g^s_1$. For
the two quark system in the lower right two by two block matrix
we can use the results of ~\cite{BR}. The remainig matrix elements describe
transitions from quark to gluon states, they are taken from the
previous section.
The second term on the rhs
of (\ref{3.8}) belongs to the gluon bremsstrahlung diagrams.
The matrix $G_0$ has the form:
\begin{equation} \label{2.32}
G_0\ =\ \left(\begin{array}{ccc}
C_A & 0 & 0\\
0 & C_F & 0\\
0 & 0 &C_F \end{array} \right).
\end{equation}
In analogy
to the matrix $F_0$ which carries color zero in the t-channel
we define the matrix $F_8$
of color octet amplitudes. Due to the antisymmetric color structure
the elements of $F_8$ are even signature amplitudes. They satisfy evolution
equations ~\cite{KL} similar to (\ref{3.8}):
\begin{equation} \label{40}
F_8\ =\ \frac{g^2}\omega\ M_8+\frac{g^2C_A}{8\pi^2\omega}\
\frac d{d\omega}\ F_8(\omega) +\frac1{8\pi^2\omega}\ F_8(\omega)^2.
\end{equation}
The color factor $C_A$ in front of the second term on the rhs is the
analogue of the matrix $G_0$ in (\ref{3.8}). The difference between
$C_A$ in (\ref{40}) and $G_0$ in (\ref{3.8}) is due to the fact that
for the positive signature amplitude $F_8$ the sum of the two
bremsstrahlungs diagrams is independent of the
type of the incoming partons, and the matrix of color factors $G_8$
becomes $C_A$ times the unit matrix.
Finally, the matrix $M_8$ is the analogue of the matrix $M_0$ in
(\ref{3.8}), but for color octet quantum numbers in the $t$-channel instead
of color singlet. It reads:
\begin{equation} \label{2.29}
M_8\ =\ \left(\begin{array}{ccc}
2C_A & -T_F & 0\\
C_A & -1/2N_c & 0\\
0 & 1/2N_c & -1/2N_C \end{array} \right).
\end{equation}

\section{Solution of the evolution equation}

The solution for $g^s_\perp$, that is
our vector $T$ (eq.(34,35)), is obtained by solving Eq.(\ref{40}) for
$F_8$, then Eq.(\ref{3.8}) for $F_0$ and Eq.(\ref{3.7}) for $R$ (the
latter makes use of the Born approximation, $R_B$, as initial condition
for $R$), and finally inserting $R$ into (\ref{3.2}). The final answer
for the three component vector $T$ will have the form:
\begin{equation} \label{4.1}
T\ =\ \int\frac{d\omega}{2\pi i}\left(\frac1x\right)^\omega
\xi(\omega)\left(\frac{Q^2}{\mu^2}\right)^{F_0/8\pi^2} \frac1{\omega
-F_0/8\pi^2}\ R_B\ .
\end{equation}
where
\begin{equation}
R_B\ =\ \left(\begin{array}{c}
0 \\ 2e_q^2 \\ 2 e_q^2 \end{array} \right)
\end{equation}
Let us go through these steps in somewhat more detail.

We begin with the equation for $F_8$. We first diagonalize the Born
term, i.e. the matrix $M_8$. Thanks to the
vanishing of the off-diagonal elements $(M_8)_{sg}=(M_8)_{sk}=0$ the two
(largest) eigenvalues $\lambda^{\pm}$ coincide with the analogue
eigenvalues for the DL evolution of the singlet function $g^s_1$.
They are given by:
\begin{equation} \label{4.2}
\lambda^{(\pm)}_8\ =\ \frac{2C_A-1/2N}2 \pm\frac12
\sqrt{(2C_A+1/2N)^2-4C_AT_F}
\end{equation}
and
\begin{equation} \label{4.2a}
\lambda^{(s)}_8\ =\ -\ \frac{1}{2N_c}.
\end{equation}
Let $e^{(+)},e^{(-)}$ and $e^{(s)}$ denote the eigenvectors of $M_8$:
\begin{equation} \label{4.11}
e^{(+)}=\left(\begin{array}{l} 1\\ x^{(+)}\\ y^{(+)} \end{array}
\right), \quad x^{(+)}=\ \frac{\lambda^{(+)}-M_{11}}{M_{12}}\ , \quad
y^{(+)}=\ \frac{\lambda^{(s)}x^{(+)}}{\lambda^{(s)}-\lambda^{(+)}}
\end{equation}
and
\begin{equation} \label{4.12}
e^{(-)}=\left(\begin{array}{l}  x^{(-)}\\ 1\\ y^{(-)} \end{array}
\right), \quad x^{(-)}=\ \frac{\lambda^{(-)}-M_{22}}{M_{21}}\ , \quad
y^{(-)}=\ \frac{\lambda^{(s)} }{\lambda^{(s)}-\lambda^{(-)} }\ .
\end{equation}
The last eigenvector corresponds to the pure quark eigenvalue
$\lambda=\lambda_8^{(s)}$ and reads:
\begin{equation} \label{4.12a}
e^{(s)}\ =\ \left( \begin{array}{l} 0\\ 0\\ 1 \end{array} \right) .
\end{equation}
Denote by $E_8=(e^{(+)}, e^{(-)}, e^{(s)})$ the matrix composed of these
eigenvectors.
For the diagonalization we also need its inverse $E_8^{-1}$:
\begin{equation} \label{4.13}
E^{-1}=\ \frac1{1-x^{(+)}x^{(-)}}\left(\begin{array}{ccc}
1 & -x^{(-)} & 0\\
-x^{(+)} & 1 & 0\\
b^{(+)} & b^{(-)} & 1-x^{(+)}x^{(-)} \end{array} \right),
\end{equation}
where
\begin{equation} \label{4.13a}
b^{(+)}=\ y^{(-)}x^{(+)}-y^{(+)}\ \mbox{  and  }\ b^{(-)}=\
y^{(+)}x^{(-)}-y^{(-)} .
\end{equation}
Then we can diagonalize the matrix $M_8$:
\begin{equation} \label{4.3}
M_8\ =\ E_8\widehat M_8E^{-1}_8\ ,
\end{equation}
where $\widehat M_8=\rm diag(\lambda^+_8,\lambda^-_8,\lambda_8^{(s)})$.
Consequently, eq.(\ref{40}) becomes diagonal if we transform to
$\widehat F_8$:
\begin{equation} \label{4.4}
F_8\ =\ E_8\widehat F_8E^{-1}_8\ .
\end{equation}
Using the ansatz $\widehat F^i_8=\ N_cg^2\ \frac\partial{\partial\omega}
\ln u^i(\omega)$
for the nonlinear Riccati equation (43), one finds second order linear
differential equations for the $u^i$. Their solutions are given by parabolic
cylinder functions. As a results we find for the components of
$\widehat F_8$:
\begin{equation} \label{4.5}
\widehat F^i_8\ =\ N_c g^2\ \frac\partial{\partial\omega}\ln e^{z^2/4}
D_{p_i}(z)\ ,
\end{equation}
where $D_p$ denotes the parabolic cylinder function with
\begin{equation} \label{4.6}
p_i\ =\ \frac{\lambda_8^{(i)}}{N_c}\ , \qquad (i=+,-,s)
\end{equation}
and
\begin{equation} \label{4.7}
z\ =\ \frac\omega{\omega_0}\ ,\qquad \omega_0=\ \sqrt{N_c g^2/8\pi^2}\ .
\end{equation}

With this solution for $F_8$ we return to the evolution
equation(\ref{3.8}) for $F_0$ which is solved by the (matrix-valued)
expression:
\begin{equation} \label{4.8}
\frac1{4\pi^2\omega}\ F_0=\ 1-\sqrt{1-\frac{g^2}{2(\pi\omega)^2}\,M_0
+\frac{g^2}{4\pi(\pi\omega)^3}\,G_0F_8}\ .
\end{equation}
We will need the right-most singularity in the
$\omega$ plane of the matrix $F_0$.
Similar to the case of $g_1$, this singularity
is due to the vanishing of the square root in (\ref{4.8}), i.e. we need
to determine the zeroes of the eigenvalues of the matrix under the
square root. The diagonalization of this matrix is done in exactly the
same way as for $M_8$: again, the matrixelements `$sg$' and `$sk$' are zero,
and the two largest eigenvalues are the same as for the singlet function
$g_1$. The diagonalization is done through the matrix $E_0$ which consists of
the corresponding three eigenvectors $e^{(+)}_0$, $e^{(-)}_0$ and $e^{(s)}_0$
(analogous to (50) -(52)).
As it was discussed in \cite{BER2}, the largest eigenvalue
$\lambda^{(+)}$ is not too far from the value obtained in the pure
gluonic case, neglecting the quark contribution.
The accurate values have to be found from a numerical computation of
the parabolic cylinder functions. For the case of $n_F=4$ we found in
\cite{BER2} the rightmost singularity $\omega_s=3.45\omega_0$ (the pure
gluonic case would have given 3.66 instead of 3.45, or even 4 if one
neglects the non-ladder DL contribution). With $\omega_0$ from
(\ref{4.7}) and $\alpha_s=0.18$ we find for the leading term in
(\ref{4.1})
\begin{equation} \label{4.20}
\omega_s\ = \ 1.01\ .
\end{equation}
Two other singularities correspond (for $n_F=4$) to $z^{(-)}=1.81$ and
$z^s=1.39$.

Having found the matrix $F_0$, we return to (40) and find the
matrix $R$:  \begin{equation} R=\left(\frac{Q^2}{\mu^2}
\right)^{\frac{F_0}{8\pi^2} - \omega} \hat{R}(\omega) \end{equation} where
the matrix valued function $\hat{R}(\omega)$ has to be determined from the
initial conditions of the evolution equation, from the energy dependence of
$R$ at the point $y=0$, i.e. at $Q^2=\mu^2$.  At this point the matrix $R$
satisfies another evolution equation (which has been discussed in detail in
~\cite{BER1,EMR} (after eq.(3.14)), and its solution has the form:
\begin{equation}
\hat{R}(\omega) = E_0 \frac{1}{\omega - \hat{F}_0/8 \pi^2} E_0^{-1} R_B.
\end{equation}
 Retaining in $\widehat F_0$ only the leading upper component we find
for the behavior of $R$ near the square root branch point at
$\omega=\omega_s$:
\begin{equation} \label{4.21}
R(\omega,y)\ \sim\ \frac{-2e^2_q\cdot x^{(-)} }{1-x^{(+)}x^{(-)} }
\left(\begin{array}{l}
1 \\ x^{(+)} \\ y^{(+)} \end{array} \right)
\frac2{\omega_s} \left(\frac{Q^2}{\mu^2}\right)^{\omega_s/2}
\end{equation}
with $x^{(+)}=0.29$, $x^{(-)}=0.43$, $y^{(+)}=-0.052$, and $y^{(-)}=-1.40$
Without the non-ladder contribution (i.e. putting $(F_8=0)$)
we would have $x^{(+)}=0.28$ and $y^{(+)}=-0.039$.

Finally, to obtain the structure function $g^s_2$ we have to subtract
the known function $g^s_1$ \cite{BER2} ($g_2=g_\perp-g_1$), and to add the
positive signature part $(-g^L/2)$ given by the same pure ladder DL
evolution as in the nonsinglet case \cite{BR}:
\begin{equation}
g_\perp=g_1+g_2 \ =\  \int\frac{d\omega}{2\pi i}\left(\frac1x\right)^\omega
\xi(\omega) \left(\frac{Q^2}{\mu^2} \right)^{\frac{\hat{F}_0}{8\pi^2}}
E_0 \frac{1}{\omega - \hat{F}_0/8 \pi^2} E_0^{-1} R_B -g^L/2
\end{equation}

\section{Conclusions}

In this paper we have studied the small-$x$ behavior of the singlet
polarized structure function $g_2$, using the double logarithmic
approximation. Our analysis is based upon ladder-type Feynman diagrams,
but in order to make our DLA study complete, we had to take into account
also the bremsstrahlung of soft gluons.
We have shown that, at small $x$, the leading contribution to
$g_2^s(x,Q^2)$ is given by the convolution of the
nonsinglet spin
dependent distribution (corresponding to the quark density matrix
$(s^\perp k)\gamma_5\hat k$) with the singlet function $g_1^s$.
As a novel feature, $g_2$ contains pieces with
different signature, both for the nonsinglet and for the singlet case.
At small $x$, we therefore have to decompose the Feyman diagrams,
in order to seperate even and odd signature.
% come from the odd-signature part
% of the amplitude.
The small-$x$ asymptotic behaviour of the structure function $g^s_2$ 
comes from the odd-signature part of the amplitude and is
driven by the asymptotics of the singlet spin dependent function 
$g^s_1$. At $1/x\to\infty$ we have found
\begin{equation} \label{29}
g_2\ \propto\ \left( \frac{1}{x}\right)^{\omega_s} \left(\frac{Q^2}{\mu^2}
\right)^{\gamma}\ ,
\end{equation}
where the anomalous dimension $\gamma$ has the value
\begin{equation}
\gamma=\omega_s/2
\end{equation}
and
\begin{equation}
\omega_s\simeq3.45\sqrt{\alpha_sN_c/2\pi}
\end{equation}
(the latter result holds for $n_F=4$).
In contrast to the unpolarized case, at small-$x$ the gluons mix with the
quarks. As in the singlet polarized structure function $g_1^s$, the dominant
contribution comes from the $t$-channel two gluon states. This is mainly
due to the color charge of the gluon which is much
larger than that of the quark.
%\footnote{Pure gluonic evolution gives $\omega_s=
%3.66\sqrt{\alpha_sN_c/2\pi}$; %without the non-ladder DL contribution 
%the pure gluonic case corresponds to
%$\omega_s=4\sqrt{\alpha_cN_c/2\pi}$ \cite{BER2}.}.
Numerically the value
of $\omega_s$ is rather large: $\omega_s\simeq1.01$ for
$\alpha_s=0.18$. This value of $\omega_s$ is
considerably larger than the value of corresponding "intercept" of the
nonsinglet function $g^{n.s.}_2$. It is also larger than the even
signature part $-g^L/2$ of the singlet $g_2^s$,
where
$\omega_{n.s.} =\omega_0 =\sqrt{(2C_F\alpha_s)/\pi}$ (the ratio of the two
values, $\omega_s/\omega_0$ is about 2.6). Therefore the asymptotic
behaviour of the spin dependent singlet structure function
$g^s_2(x,Q^2)$ is almost completely driven by the DL behavior of the singlet
$g^s_1(x,Q^2)$ distribution.

From the formal point of view,
the double logarithmic approximation (DL) that we have been using here
(i.e.the neglect of nonlogarithmic contributions) can be justified
only for the case of a very small QCD coupling $\alpha_s\ll1$, in which case
also $\omega_s$ is small. The experience from the LO and NLO BFKL 
calculations supports the belief that, whenever the leading
approximation turns out to be large, one has to expect also large NLO
corrections. Thus one should be careful in using the DL results for the
singlet structure function for the numerical applications.
Nevertheless, our findings indicate that the singlet
spin dependent structure function $g^s_2$ grows steeply at $x\to0$.

\end{document}